\documentclass[useAMS,usegraphicx,usenatbib]{mn2e}

\title[QPO states of 4U 1630--47]{The QPO states of 4U 1630--47}
\author[M Choudhury et al.]{Manojendu Choudhury$^{1}$\thanks{E-mail:manojendu@cbs.ac.in (MC)}
\newauthor Nilay Bhatt,$^{2}$  Subir Bhattacharyya$^{2}$
\\
$^{1}$UM-DAE Centre for Excellence in Basic Sciences, Vidyanagri Campus, Mumbai-400098, India\\
$^{2}$Astrophysical Sciences Division, Bhabha Atomic Research Centre, Mumbai-400088, India
}

\begin{document}

\date{Accepted 2014 December 23. Received 2014 December 23; in original form 2013 September 19
 }

\pagerange{\pageref{firstpage}--\pageref{lastpage}} \pubyear{2002}

\maketitle

\label{firstpage}

\begin{abstract}

Among the transient black hole binary systems, 4U 1630--47 is one of the most active
sources exhibiting outbursts every few hundred days, with every outburst lasting 
typically around hundred days. During the 2002--2004 outburst the appearance of quasi-periodic oscillations (QPO) coincide with the onset of anomalous state. There are two distinct QPO states: namely the single QPO state with one QPO and the twin QPO state with two QPOs not related harmonically. The spectral features of this state are corroborated by a previous outburst in 1998. The evolution of the inner disc temperature and the inner disc radius suggest the possible onset of geometrically thicker, ’slim’ disc as a possible explanation of the energy spectral features. The other possibilities involving the Comptonizing cloud also exist that may explain the anomalous state. The two different QPO states exhibit different spectral features, and we provide a possible empirical physical scenario from the observation of the evolution of the spectral features.
\end{abstract}


\section{Introduction}
Since its first observations from \textit{Uhuru} followed by \textit{Ariel V} in the early seventies, the X-ray binary source 4U 1630--47 \citep{jones76apj} has exhibited regular outbursts with a gap of, typically, few hundred days \citep{priedhorsky86apss}. The nature as well as the durations of these outbursts can vary greatly \citep[see for example][]{abe05pasj,tomsick05apj}, hence the study of spectral and the timing properties of the source can play a very important role in the fundamental understanding of the accretion process for the transient X-ray binary systems in general.

Neither the binary orbital period nor the nature or mass of the companion of 4U 1630--47 is known \citep[see][for a possible identification of the infrared counterpart]{augusteijn01aa}. Nevertheless, based on the various observations in the X-rays from \textit{Einstein} \citep{parmar97aa}, \textit{EXOSAT} \citep{parmar86apj,kuulkers97apjl}, \textit{Ginga} \citep{parmar97aa}, BATSE/\textit{CGRO} \citep{bloser96aas}, \textit{ROSAT} \citep{parmar95apjl}, \textit{Beppo-SAX} \citep{oosterbroek98aa}, \textit{ASCA} \citep{parmar97aa} and \textit{RXTE}\citep{kuulkers98apj,tomsick98apj,hjellming99apj,tomsick00apj,
dieters00apj,trudolyubov01mnras,kalemci04apj}, \citet{mcclintock06} classified the source as a `category A' black hole candidate, supporting the earlier classification by \citet{tanaka95}. 

Although the earlier observations of this source, including those till \textit{ROSAT} studied mainly the outburst periodicity and the canonical state transitions, the power density spectra (PDS) obtained from the lightcurves as observed by \textit{EXOSAT} did suggest some anomalous behaviour \citep{kuulkers97apjl}. Thereafter, the spectral observations from \textit{Beppo-SAX} \citep{oosterbroek98aa} suggested the inadequacy of the optically thick and geometrically thin disc models under certain circumstances. The \textit{RXTE}/PCA observations of the outbursts in the early \textit{RXTE} era, viz. 1998 outburst \citep{hjellming99apj} brought the quasi-periodic oscillations (QPO) observed in this source to the fore-front of study and classification \citep{dieters00apj,tomsick00apj,trudolyubov01mnras}. 

\begin{figure}
\includegraphics[width=0.4\textwidth, angle=-90]{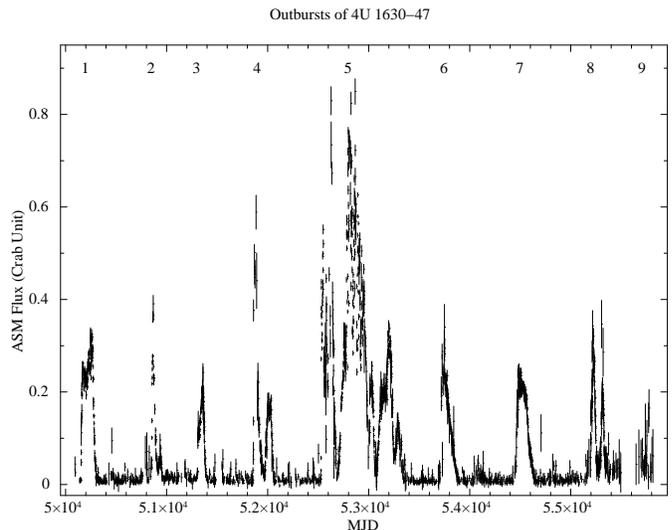}
\vspace{-4.0mm}
 \caption{All the outbursts of the source 4U 1630--47 covered by the \textit{ASM} during 1996-2011. The source shows a total of 9 outbursts with a few outbursts exhibiting multiple flares, the longest being the outburst 5 during 2002-2004 \citep[see][]{abe05pasj}. } 
 \label{fig_01}
\end{figure}

Among the nine outbursts exhibited by the source 4U 1630--47 during the \textit{RXTE} era (Fig.1 \ref{fig_01}), the fifth one starting from $\sim$MJD 52525 to $\sim$MJD 53344 (2002 September - 2004 December) 4U 1630--47 exhibited the longest and brightest outburst phase observed till date. \citet{abe05pasj} report the presence of three distinct spectral states including one anomalous state during the outbursts observed by \textit{RXTE} from 1996 to 2004. \citet{tomsick05apj} report about the high amplitude variability, spectral states and transitions and also state that QPOs were observed during the 2002-2004 outburst. Both \citet{abe05pasj} and \citet{tomsick05apj} suggest the presence of geometrically thick accretion i.e. slim disc to be present during some observations. \citet{tomsick05apj} also speculate that the unphysical low values of the inner accretion disc radius (as per the geometrically thin disc model) may be explained by the spectral hardening due to dominance of electron scattering in the inner region of the accretion disc giving rise to observed values of high temperature and luminosity.

Despite the extensive coverage of this source in the literature, a comprehensive correlated study of the timing features, namely the QPOs, and the spectral parameters has not been carried out. 
In this paper we report the spectral properties of the source during the 2002-2004 outburst viz-a-viz the QPO features observed. The result is corroborated by the measured parameters reported by \citet{trudolyubov01mnras} from the 1998 outburst. From the observational features of the spectral parameters and the QPO frequencies we conclude that the presence of low frequency QPOs occurs during a spectral state not classifiable by the present canonical black hole classification scheme \citep{mcclintock06} and the possibility of the existence of a slim disc appears to be a viable alternative during the peak of at least a few of the outbursts of 4U 1630--47. Other alternatives may include non-standard Computerization in the innermost region of accretion flow. Nevertheless, this source is the perhaps the only one reported so far to demonstrate significantly a different spectral state when the low-frequency QPOs appear.

\section{Data and Analysis}

\begin{figure}
\includegraphics[width=0.36\textwidth, angle=-90]{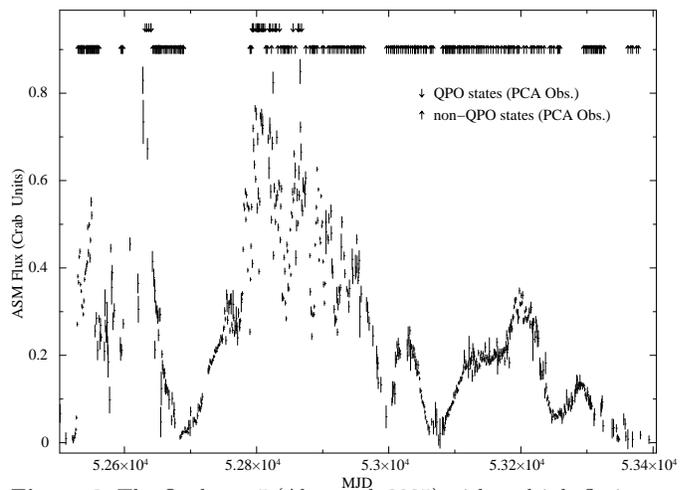}
\vspace{-4.0mm}
 \caption{The Outburst 5 \citep{abe05pasj} with multiple flaring episodes of 4U 1630--47 observed by the \textit{ASM}. The rate is measured in the units of Crab emission rate (75 counts s$^{-1}$ of \textit{ASM}). The arrows give the dates of PCA observations, the upward arrows represent observation for which no QPO is detected, and the downward arrows represent observations for which QPOs (one or more) are detected. } 
 \label{fig_02}
\end{figure}

The All Sky Monitor \citep[\textit{ASM};][]{levine96apjl} aboard the \textit{RXTE} satellite observatory \citep{bradt93aas} monitored the source continuously throughout its existence, in the 1.5 - 12 keV range. In addition the narrow field of view instrument, the Proportional Counter Array (\textit{RXTE}/PCA) has covered the source extensively, especially during the outburst phases. For the results reported in this paper, we have analysed 382 pointed observations of the \textit{RXTE}/PCA (Fig. \ref{fig_02}), including 24 observations with the telescope pointing away from the source (offset $\leq 1^{\circ}$), i.e. high-offset observations. The lightcurves were extracted from the event mode data for the 3-30 keV energy band and the complete band, at a time resolution of $2^{-8}$ s. The energy spectra for all observations were extracted from standard 2 mode data. To maintain uniformity only PCU 2 was used for the complete analysis, as it was the only common PCU available for all observations. All the procedures of data filtering, background and deadtime corrections were strictly adhered to during the analysis process, including the cases of high-offset observations. The data reduction and analysis was carried out using \texttt{HEASOFT}\footnote{http://heasarc.gsfc.nasa.gov/docs/software/lheasoft/}, which consists of (chiefly) \texttt{FTOOLS}, \texttt{XRONOS} and \texttt{XSPEC} \citep{arnaud96aspc}. The power density spectrum (PDS) were obtained from the event mode data for the complete energy band covered by the \textit{RXTE}/PCA using the \texttt{IDL} based software package \texttt{GHATS}\footnote{package developed by T.M. Belloni at INAF - Osservatorio Astronomico di Brera}.

The PCA spectra were fitted for all the observations with a single three-component model consisting of multicoloured disc blackbody dominating at lower energies (\emph{diskbb}), a \emph{power law} at higher energies and an isotropic reflection component (\emph{reflect}) of the \emph{power law} emission from the accretion disc (along with the interstellar absorption), as well as a two component model without the reflection component.
Statistically good fits were obtained using the three component model for almost all the observations with the reduced $\chi_{\nu}^2 \leq 1.20$ (with a systematic error $\leq 1\%$) while fitting the model to PCA spectra in the 3-30 keV band using the $\chi^2$-minimization algorithm of the \texttt{XSPEC} package. For the complete set of observations considered, the best-fitting values of disc temperature $T_{in}$ varies in the region 0.7-3.2 keV, while the \emph{power law} index $\Gamma$ varies from 1.5 to a very steep value of 4.8. Although the possibility of any contamination from other sources in the energy range of our analysis is minimal despite the location of the source being in a crowded region  \citep{tomsick05apj}, we report the spectral parameters for the observations with 3-30 keV flux $> 1.2\times 10^{-9} erg\, cm^{-2} s^{-1}$, just to ensure the removal of any effect on the spectral features due to contamination when the intrinsic flux from the source is weaker.

The reflection component (\emph{reflect}) was used to account for the broad excess around $8-10$ keV for some observations in order to better constrain the inner disc temperature $T_{in}$ of the \emph{diskbb} model \citep{makishima86apj} in \texttt{XSPEC}. Although this reflection component did not constrain the high values of $T_{in}$ for certain observations, the $\chi^2$ value was better constrained using this component for such observations, hence for the sake of consistency  all the results reported in this paper are with the reflection component for all observations. The lower limit of the column density parameter ($N_H$) was restricted to $6\times10^{22} cm^{-2}$, which is the lowest measured value for 4U 1630--47 \citep{parmar97aa}. Following the study of X-ray dips in the source \citep{kuulkers98apj} the value of $60^o$ was used for the inclination angle, although, admittedly, there is huge uncertainty in this value as none of the binary parameters nor the companion star properties are ascertained. Similarly, the distance of the source was assumed to be $\sim 10 kpc$. Consequently, the conclusions of this paper remain a qualitative statement based on the long term trends of the parameters as observed in the source. In case the actual distance or the angle of inclination is greater then, in either case, the value of $R_{in}$ will be systematically greater.

For certain high inner disc temperature $T_{in}$, the corresponding value of inner disc radius $R_{in}$ is not always physical (Fig. \ref{fig_05}, Section 5.1). For better constraint on the best-fitting values of these parameters, an attempt was made to use an empirical Comptonization model \emph{simpl} which functions as a convolution with the \emph{diskbb} component converting a fraction of the input seed photons into a \emph{power law} \citep{steiner09pasp}. This model failed to fit the data statistically and the addition of additional Comptonizing component \citep[namely \emph{comptt}, ][]{titarchuk94apj} was needed to obtain an acceptable $\chi^2$ value. But even this model failed to constrain the values of $T_{in}$ and $R_{in}$ within physically acceptable limits. Also, a model consisting of only \emph{diskbb} and \emph{comptt} failed to constrain the values of $T_{in}$ and $R_{in}$. Hence these models were discarded in favour of the model described in the previous paragraphs to compare the current observations with the standard classification.

In addition to using the \texttt{GHATS}, quite a few PDS (especially for the large number of observations without any QPO) were extracted using the \texttt{ISIS}\footnote{http://space.mit.edu/CXC/ISIS} \citep{houck00aspc,houck02} package with an add-on module called \texttt{SITAR}\footnote{http://space.mit.edu/CXC/analysis/SITAR}. Of all the pointed observations analysed pertaining to the 2002-2004 outburst, statistically significant QPO was found for only 30 observations. All the PDS were fitted with single model consisting of peaked noise components \citep{psaltis99apj} comprising of a zero-frequency centred Lorentzian (zfc) and four Lorentzians \citep{nowak00mnras,belloni02apj}, as defined below:
\[ \textrm{zfc} = \frac{R_0}{1+\nu/\nu_0} \] and \[ \textrm{Lor} = \nabla^{-1}\frac{R^2Q\nu_0}{\nu_{0}^{2}+Q^2(\nu-\nu_0)^2}  \]

\noindent where in the Lorentzian component, $\nu_0$ is the  resonant frequency of the Lorentzian, $Q$($=\nu_0/\Delta\nu_0$, where $\Delta\nu_0$ is the FWHM of the Lorentzian centred at $\nu_0$) is the quality factor and R is the normalization of the Lorentzian, such that the rms variability, rms =$ R[0.5-tan^{-1}(-Q)/\nabla]^{-1/2}$ (i.e. rms = $R$ as $Q\rightarrow 0$). 
The minimum criterion for the existence a QPO is that the corresponding Lorentzian should have Q$>$1, while the preferable condition is that Q$>$2 and this latter condition was satisfied for most of the observations.  
On many of the 30 observations with QPOs there exist multiple Lorentzians with $Q\sim2$ or greater, but the centroid frequencies of the Lorentzians rule out the presence of harmonics or sub harmonics. The PDS was converted into a FITS file similar to energy spectra along with the creation of a dummy response matrix using the \texttt{GHATS} package and the $\chi^2$ minimization technique of the \texttt{XSPEC} was utilized to obtain the QPO parameters.

\begin{figure}
\includegraphics[width=0.5\textwidth]{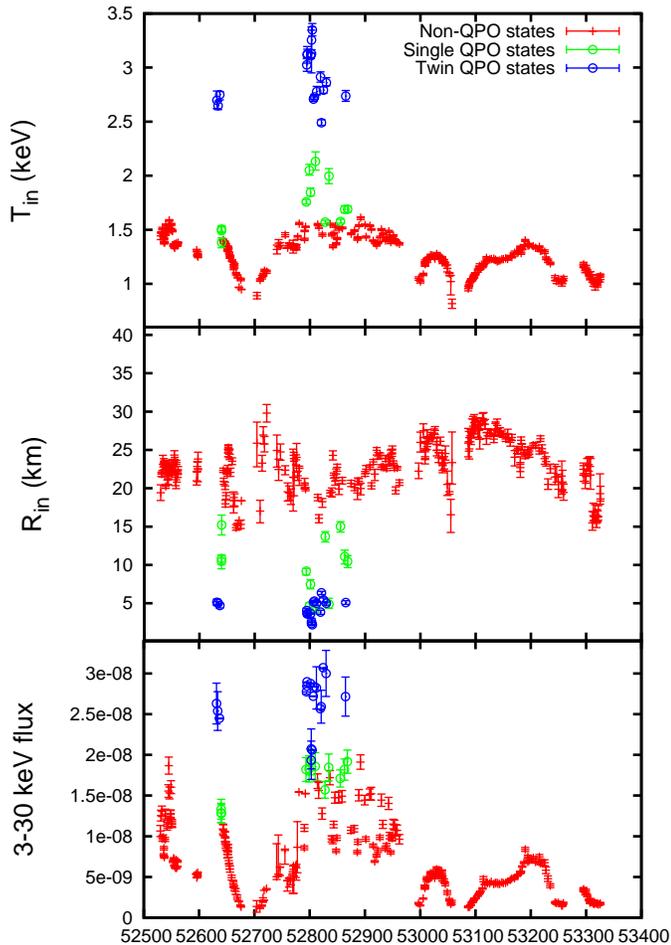}
\vspace{-5.5mm}
 \caption{The outburst no. 5 \citep{abe05pasj} of 4U 1630--47 observed by the PCA. Bottom panel: 3-30 keV flux ($\times10^{-8}erg\, cm^{-2} s^{-1}$) as a function of days for which the PCA observations exist. Middle panel: the inner radius $R_{in}$ of the accretion disc obtained from the \emph{diskbb} model \citep{makishima86apj}. Top panel: the inner disc temperature $T_{in}$ as per the \emph{diskbb} model. For all panels: the red marks indicate observations for which no QPO is present, the green and blue marks indicate observations with single and twin QPOs, respectively.} 
 \label{fig_03}
\end{figure}

\section{The Extended Outburst of 4U 1630--47}
During the \textit{RXTE} era the source went into outburst phase multiple times, where some outbursts consisted of more than one flaring episode, including the 2002-2004 outburst which is referred to as outburst 5 \citep[following the convention of][]{abe05pasj}. A similar long outburst phase with more than one flaring episode probably occurred during 1988-1991 (total duration $\sim 2.4$ yr) as observed by \textit{Ginga}, when the source flux reached a peak value of $\sim 0.6$ Crab \citep{kuulkers97mnras}. Furthermore, the source perhaps exhibited very bright outburst in 1977 \citep{chen97apj} but the duration of the outburst is not ascertained due to the sparse coverage of the source in that era of X-ray observatories. In comparison \textit{RXTE} provides a continuous monitoring, via the ASM, as well as wide band spectral coverage coupled with high resolution timing studies, via the PCA, from 1996 to 2011.
Other than the 2002-2004 phase, the duration of the other outbursts is typically $\sim 100$ d (give and take few tens of days). Also, these typical short duration outbursts emit at a peak flux of $\sim 0.4-0.5$ Crab \citep{abe05pasj,tomsick05apj} (see Fig. \ref{fig_01}).

The long outburst 5 comprised of five individual flaring episodes, each immediately following the previous one with a gap of few days (Fig. \ref{fig_02}), with the individual flaring episodes being referred to alphabetically A, B etc. The duration of flare 5A was from MJD 52525 to 52675, the flare 5B was during MJD 52686 to 52996, the flare 5C was during MJD 53004 to 53061, the flare 5D was during MJD 53077 to 53251, the flare 5E was during MJD 53252 to 53344. The flares 5A and 5B were evidently different from the other three flares as per the \textit{ASM} data. The peak flux of these two flares reach a value $\sim 0.6$ and $\sim 0.8$ Crab, whereas the other three flares do not get brighter than $\sim 0.3 - 0.5$ Crab. Also, the peak phase of the flares 5A and 5B show large and fast fluctuations, whereas the other three show more stable behaviour.
The arrows of Fig. \ref{fig_02} are the days for which the pointed observation by PCA is available in the archives. The downward arrows indicate observations for which QPOs have been observed, while the upward arrows indicate observations with QPOs. 

In Fig. \ref{fig_03} bottom panel the 3-30 keV flux ($\times10^{-8} erg \, s^{-1} cm^{-2}$) measured by the PCA during outburst 5 is plotted. The evolution of the inner disc radius and temperature, $R_{in}$ and $T_{in}$, as obtained from the model component \emph{diskbb}, are plotted in the middle and top panels, respectively. The data points of Fig. \ref{fig_03} are segregated as non-QPO (red mark) and QPO (green and blue marks) state. 

All the QPOs occur during the flares 5A and 5B, and that too only during the peak phase when the total flux as well as the spectral parameters show fast and drastic fluctuations. These fluctuations occur when, during the rising phase of the flare, the flux increases beyond $\sim 0.3 - 0.4$ Crab flux. The fluctuations continue until there is a definite decay of the flare and the flux comes down below that same value, viz $\sim 0.4$ Crab flux. In between, during the peak phase fluctuations, the flux can reach a value less than $\sim 0.4$ Crab, and the parameter values of the inner disc is not discernible from the canonical black hole states. Hence, a more rigorous insight into the data is provided below to ascertain the behaviour of the accretion disc of the low flux phases during the peak phase fluctuations. In this paper we have concerned our analyses only on averaged measurements of the different observational pointings.

\section{Different classes of QPO\small{s}}

\begin{figure}
\includegraphics[width=0.5\textwidth]{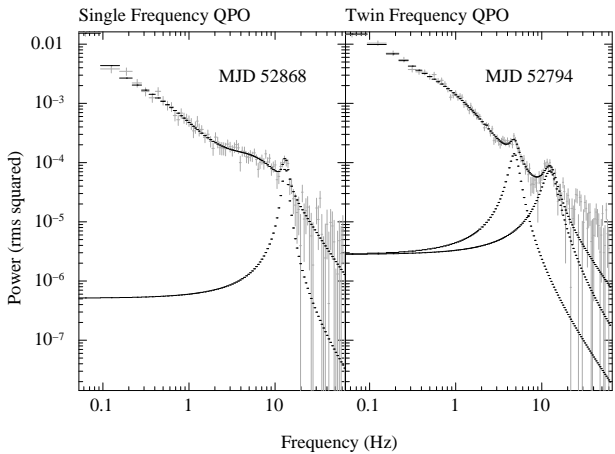}
\vspace{-11.5mm}
 \caption{Examples of the single and twin QPO states.} 
 \label{fig_04}
\end{figure}

The observed PDS with QPOs are predominantly divided into two distinct classes, one with one QPO with the centroid frequency in the region of 11-14 Hz (henceforth referred to as `single QPO state'), and the other with two non-harmonically related QPOs with the centroid of the lower frequency lying in the region of 4-6 Hz while the higher frequency can vary in the region of 7-18 Hz (henceforth referred to as `twin QPO state'). A sample each of the single QPO and the twin QPO is given in Fig. \ref{fig_04}. In addition, another class of lowest frequency QPO with a periodicity around 1 Hz is observed on a few occasions, and only once it appears alone in the absence of the other two classes of QPOs. Due to lack of statistically significant data, we will not include the lowest frequency QPO in our analysis. The details of the QPO parameters are given in table \ref{tab_01}.  The significance of the QPO was obtained from the errors calculated from the best-fitting statistics for the normalization parameter of the corresponding Lorentzian function.

For the single QPO states, the peak rms as well as the significance is relatively greater for lower frequencies (the peak rms$\;\geq\!\!1\%$ for frequencies $<\!12$ Hz and the significance is highest for the lowest frequencies - table \ref{tab_01}). For the twin QPO states both the peak rms as well as the significance shows large variation, but in general the peak rms of the lower frequency is equivalent or more than that of the higher frequency. The total flux measured from the source is more for twin QPO state except for the occasions when the higher frequency is in the range of 7 Hz (table \ref{tab_01}).

\begin{table*}
\caption{The details of the observed QPO. The MJD (with the fractional part) denotes the start of the observation (there are occasions of multiple observations starting on the same day). The significance of the QPO is obtained from the best-fitting parameters of the PDS, see the text for details. }
\label{tab_01}
\begin{tabular}{@{}lccccccc}
\hline \hline
$ $ & MJD & Obs. Id. & QPO freq. & FWHM	& rms-peak(\%) & sig. & PCU2 count rate	\\
\cline{2-8}
Single QPO & & & & & & & \\
& 52639.84286 & 70417-01-09-01 & $11.79^{\pm0.25}$ & $2.39^{\pm1.09}$ & 1 & 2 & $1254^{\pm81}$ \\
& 52639.92536 & 70417-01-09-02 & $11.81^{\pm0.43}$ & $3.71^{\pm1.49}$ & 1.1 & 2 & $1184^{\pm199}$\\
& 52640.23910 & 70417-01-09-03 & $12.36^{\pm0.54}$ & $4.41^{\pm1.64}$ & 0.8 & 4 & $1278^{\pm46}$ \\
& 52793.38442 & 80117-01-02-01 & $12.53^{\pm0.12}$ & $2.40^{\pm0.46}$ & 1 & 5 & $1720^{\pm223}$ \\
& 52798.96622 & 80117-01-04-00 & $11.35^{\pm0.03}$ & $1.56^{\pm0.11}$ & 1.6 & 16 & $1757^{\pm194}$ \\
& 52800.84348 & 80117-01-02-04 & $12.46^{\pm0.13}$ & $1.63^{\pm0.42}$ & 1 & 5 & $1804^{\pm213}$ \\
& 52810.09160 & 80117-01-08-00 & $11.46^{\pm0.04}$ & $0.94^{\pm0.11}$ & 1.9 & 12 & $1737^{\pm127}$ \\
& 52827.72771 & 80117-01-11-01 & $13.28^{\pm0.25}$ & $1.14^{\pm0.61}$ & 0.6 & 2 & $1629^{\pm149}$ \\
& 52834.50147 & 80117-01-12-01 & $11.60^{\pm0.06}$ & $1.18^{\pm0.21}$ & 1.7 & 7 & $1689^{\pm259}$ \\
& 52862.84906 & 80117-01-16-00 & $13.30^{\pm0.23}$ & $1.95^{\pm0.81}$ & 0.9 & 3 & $1791^{\pm153}$ \\
& 52868.08409 & 80117-01-17-00 & $13.10^{\pm0.18}$ & $2.07^{\pm0.67}$ & 0.9 & 3 & $1865^{\pm170}$ \\

\cline{2-8}
Twin QPO & & & & & & & \\
& 52631.49685 & 70417-01-08-00 & $4.83^{\pm0.11}$ & $0.91^{\pm0.32}$ & 0.8 & 4 & $2393^{\pm402}$ \\
&  &  & $13.27^{\pm0.34}$ & $5.89^{\pm1.05}$ & 0.7 & 8 &  \\
\cline{4-7}
& 52633.66694 & 70417-01-08-01 & $5.14^{\pm0.14}$ & $0.76^{\pm0.47}$ & 0.6 & $<\!2$ & $2214^{\pm304}$ \\
&  &  & $13.22^{\pm0.36}$ & $5.60^{\pm2.11}$ & 0.6 & 2 &  \\
\cline{4-7}
& 52636.79921 & 70417-01-09-00 & $4.62^{\pm0.16}$ & $2.14^{\pm0.52}$ & 0.5 & 5 & $2164^{\pm217}$ \\
&  &  & $12.54^{\pm0.16}$ & $5.24^{\pm0.52}$ & 0.7 & 13 & \\
\cline{4-7}
& 52794.23388 & 80117-01-02-02G & $4.71^{\pm0.05}$ & $1.90^{\pm0.17}$ & 1.1 & 12 & $2366^{\pm224}$ \\
&  &  & $12.24^{\pm0.07}$ & $3.44^{\pm0.20}$ & 1 & 23 &  \\
\cline{4-7}
& 52794.91444 & 80417-01-01-00 & $4.77^{\pm0.07}$ & $1.35^{\pm0.24}$ & 1.2 & 7 & $2410^{\pm118}$ \\
&  &  & $12.34^{\pm0.19}$ & $4.96^{\pm0.57}$ & 0.8 & 12 & \\
\cline{4-7}
& 52795.28333 & 80117-01-03-00G & $4.74^{\pm0.06}$ & $2.08^{\pm0.21}$ & 1.2 & 9 & $2424^{\pm185}$ \\
&  &  & $12.28^{\pm0.08}$ & $3.26^{\pm0.24}$ & 1 & 18 & \\
\cline{4-7}
& 52801.39948 & 80117-01-05-00 & $4.19^{\pm0.20}$ & $2.05^{\pm0.65}$ & 1.9 & 4 & $2494^{\pm145}$ \\
&  &  & $12.54^{\pm0.16}$ & $3.70^{\pm0.46}$ & 2.4 & 11 & \\
\cline{4-7}
& 52802.79627 & 80117-01-06-00 & $5.61^{\pm0.13}$ & $2.75^{\pm0.37}$ & 2.9 & 6 & $1383^{\pm193}$ \\
&  &  & $7.36^{\pm0.07}$ & $0.98^{\pm0.27}$ & 2.4 & 3 & \\
\cline{4-7}
& 52802.86045 & 80117-01-06-01 & $5.67^{\pm0.04}$ & $5.00^{\pm0.47}$ & 2.1 & 6 & $1755^{\pm274}$ \\
&  &  & $7.82^{\pm0.04}$ & $0.66^{\pm0.13}$ & 2.2 & 6 & \\
\cline{4-7}
& 52804.23360 & 80117-01-07-00 & $5.72^{\pm0.17}$ & $4.72^{\pm0.37}$ & 2.1 & 7 & $1690^{\pm303}$ \\
&  &  & $7.56^{\pm0.05}$ & $0.79^{\pm0.17}$ & 1.8 & 5 &  \\
\cline{4-7}
& 52806.54943 & 80117-01-07-01 & $4.62^{\pm0.02}$ & $0.28^{\pm0.09}$ & 1.1 & 5 & $2501^{\pm136}$ \\
&  &  & $15.15^{\pm0.30}$ & $6.29^{\pm1.35}$ & 0.6 & 4 &  \\
\cline{4-7}
& 52808.24961 & 80117-01-07-02 & $4.53^{\pm0.04}$ & $0.59^{\pm0.12}$ & 1 & 7 & $2581^{\pm112}$ \\
&  &  & $14.54^{\pm0.17}$ & $6.81^{\pm0.53}$ & 0.7 & 18 &  \\
\cline{4-7}
& 52812.26814 & 80117-01-08-01 & $4.31^{\pm0.15}$ & $0.99^{\pm0.60}$ & 0.7 & $<\!2$ & $2538^{\pm114}$ \\
&  &  & $13.71^{\pm0.24}$ & $5.11^{\pm0.79}$ & 0.8 & 7 &  \\
\cline{4-7}
& 52819.22981 & 80117-01-10-00 & $4.53^{\pm0.04}$ & $0.30^{\pm0.12}$ & 0.9 & 3 & $2275^{\pm133}$ \\
&  &  & $14.63^{\pm0.37}$ & $6.46^{\pm1.56}$ & 0.6 & 3 &  \\
\cline{4-7}
& 52820.82659 & 80117-01-10-01 & $5.11^{\pm0.05}$ & $0.32^{\pm0.13}$ & 0.9 & 3 & $2366^{\pm405}$ \\
&  &  & $17.39^{\pm0.29}$ & $3.65^{\pm4.42}$ & 0.6 & 3 &  \\
\cline{4-7}
& 52824.29111 & 80117-01-11-00G & $4.41^{\pm0.03}$ & $0.39^{\pm0.09}$ & 1.1 & 6 & $2521^{\pm159}$ \\
&  &  & $14.41^{\pm0.22}$ & $6.14^{\pm0.72}$ & 0.7 & 10 &  \\
\cline{4-7}
& 52829.55105 & 80117-01-11-02 & $4.12^{\pm0.10}$ & $0.60^{\pm0.38}$ & 0.7 & $<\!2$ & $2671^{\pm164}$ \\
&  &  & $14.36^{\pm0.22}$ & $4.18^{\pm1.03}$ & 0.8 & 4 &  \\
\cline{4-7}
& 52864.86189 & 80117-01-16-01 & $4.08^{\pm0.52}$ & $3.73^{\pm1.72}$ & 1 & 2 & $2476^{\pm131}$ \\
&  &  & $14.72^{\pm1.48}$ & $12.28^{\pm0.77}$ & 0.6 & 3 &  \\

\cline{2-8}
Misc. QPO & & & & & & & \\
& 52633.66694 & 70417-01-08-01 & $1.55^{\pm0.03}$ & $0.18^{\pm0.08}$ & 1.2 & 3 & $2214^{\pm304}$ \\
& 52802.79627 & 80117-01-06-00 & $1.39^{\pm0.09}$ & $0.86^{\pm0.29}$ & 3.1 & $<\!2$ & $1383^{\pm193}$ \\
& 52802.86045 & 80117-01-06-01 & $1.39^{\pm0.09}$ & $0.86^{\pm0.29}$ & 1.8 & $<\!2$ & $1755^{\pm274}$ \\
& 52804.23360 & 80117-01-07-00 & $1.27^{\pm0.05}$ & $1.06^{\pm0.14}$ & 2.4 & 4 & $1690^{\pm303}$ \\
& 52862.84906 & 80117-01-16-00 & $0.74^{\pm0.04}$ & $0.23^{\pm0.11}$ & 2.1 & 2 & $1791^{\pm153}$ \\
& 52855.20911 & 80117-01-15-01 & $0.84^{\pm0.05}$ & $0.35^{\pm0.15}$ & 1.8 & 3 & $1729^{\pm204}$ \\
\hline \hline
\end{tabular}

\medskip 
{\em Notes. The Miscellaneous QPOs have the lowest frequency and generally occur with other QPOs, except for MJD 52855.20911. The occurrences of single \& twin QPOs are mutually exclusive.}
\end{table*} 

In total there are 30 observations with QPO, 18 of them belong to the twin QPO state while 11 belong to the single QPO state, and once the lowest frequency QPO ($\approx 0.84 Hz$) appears alone. As per the prevalent ABC classification scheme for the QPOs \citep{casella05apj}, the single QPO may be loosely classified as class C, despite the obvious deviation of the parameters from the class C values, simply because the centroid frequency rules out the A and B class. For the twin QPO state, the higher frequency deviates from C-type classification. Eventhough the lower frequency is nearer to the B-type, the QPO parameters deviate significantly from the classification scheme.

\section{The spectral behaviour of the QPO states}
Most of the observations fall in the canonical black hole states, namely steep power law, thermal dominated and intermediate while few fall in the hard state: as classified by \citet{mcclintock06}. The anomalous states pertain to the observations with QPO. Hence, for this source, the fundamental understanding of the evolution of the accretion process needs to be constructed by segregating the spectral parameters of the QPO and non-QPO observations.

\subsection{The $R_{in}$--$T_{in}$ relation and its consequences}

\begin{figure}
\includegraphics[width=0.5\textwidth]{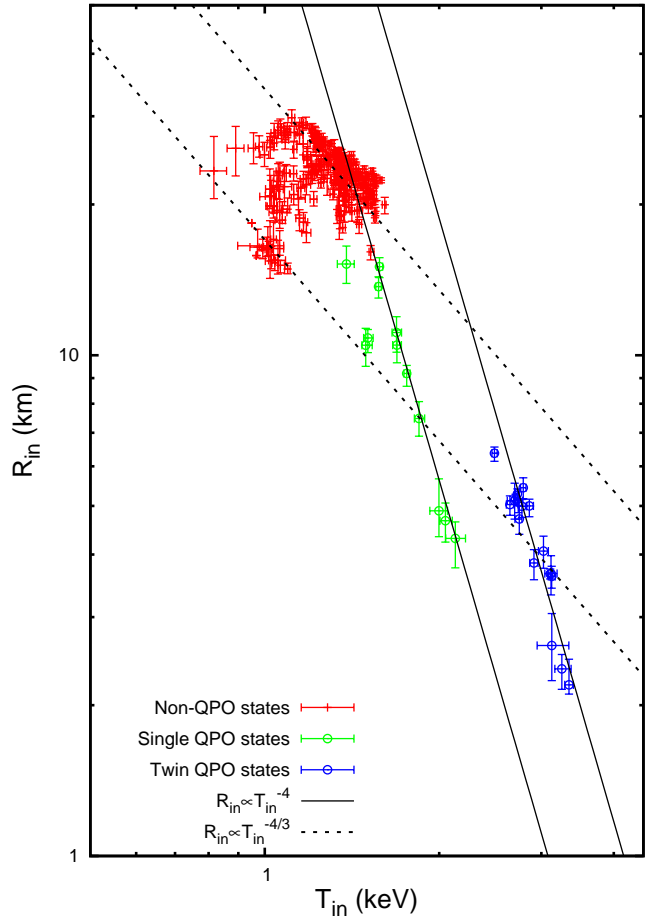}
\vspace{-8.0mm}
 \caption{The plot of inner disc radius $R_{in}$ versus inner disc temperature $T_{in}$ for all the observations pertaining to flares 5A, 5B, 5C, 5D, and 5E. The red marks represent the non-QPO states. The green marks represent the single QPO states and the blue marks represent the twin QPO states. The non-QPO observations obey the relationship $R_{in} \propto T_{in}^{-4/3}$ (geometrically thin optically thick Shakura--Sunyaev disc). The QPO observations obey the relation $R_{in} \propto T_{in}^{-4}$. } 
 \label{fig_05}
\end{figure}

The relation between the inner disc temperature, $T_{in}$ and the inner disc radius $R_{in}$ is plotted in Fig. \ref{fig_05}. Pertaining to the five flares under study, flares 5A, 5B, 5C, 5D and 5E, the $R_{in}$--$T_{in}$ plane has four branches. Two parallel branches (represented by red marks) correspond to the non-QPO observations from all the flares in the outburst. These two branches can be fitted by the empirical relation $R_{in} \propto T_{in}^{-4/3}$. This is a clear signature of Shakura--Sunyaev disc \citep{shakura73aa}. The geometrically thin and optically thick disc approximation is not violated for these observations \citep{kubota99an}. The other two parallel branches (represented by green and blue marks for QPO observations) correspond to high temperatures $T_{in}$ and low inner-radius $R_{in}$ and are best fit by a relation $R_{in} \propto T_{in}^{-4}$. This is the anomalous state which can not be modelled by a geometrically thin disc. Hence these observations provide the signature of the impending anomalous state which exists only along with the appearance of QPOs.

Nevertheless, the values of the inner disc radius for the high temperature anomalous states, as obtained from the spectral fitting, are not physically acceptable. Although it is known that different physical mechanism causing the opacity of the disc can give rise to a different spectral shape \citep{shimura95apj} which can cause an underestimation of the inner disc radius up to a factor of 5 \citep{merloni00mnras}, the most likely cause of the breakdown of the \emph{diskbb} model is the formation of geometrically thicker `slim' accretion disc \citep{abramowicz88apj}. The $R_{in} - T_{in}$ relation suggest that for a slim disc the emission at small radii can also produce a much flatter radial temperature profile due to more material in the disc \citep[][refer to fig. 4]{watarai00pasj}.

Also, the departure from the standard multi-colour disc may be attributed to effects of electron scattering and corresponding changes in radial temperature profile \citep{kubota04apj}. Both \citet{tomsick05apj} and \citet{abe05pasj} report the  deviation of the disc luminosity from the relation $L_{disc} \propto T_{in}^{4}$. While \citet{tomsick05apj} speculates about electron scattering in the inner region of accretion disc causing the high temperature and hence deviation from the $T_{in}^{4}$ relation, \citet{abe05pasj} is more convinced that this deviation is due to the slim disc. The presence of QPO suggest a rotational symmetry in the process that gives rise to the quasi-periodic phenomena, whereas higher electron scattering would result in more wide band noise, with the exception of a formation of a shock \citep{rao00aa}. The so called `anomalous' state may be classified as the QPO state, for this source, as it is with the appearance of the quasi-periodicity in the PDS the parameters pertaining to the accretion disc show a discontinuous evolution.

The two branches of the $T_{in}-R_{in}$ relation for the QPO states (Fig. \ref{fig_05}) are resolved as two distinct QPO states, with the single QPO state giving rise to the branch with lower $T_{in}$ and the twin QPO state giving rise to the parallel branch with higher $T_{in}$. The single QPO branch here includes the observation on MJD 52855.

\subsection{The \emph{powerlaw} component}
\begin{figure}
\includegraphics[width=0.53\textwidth, angle=0]{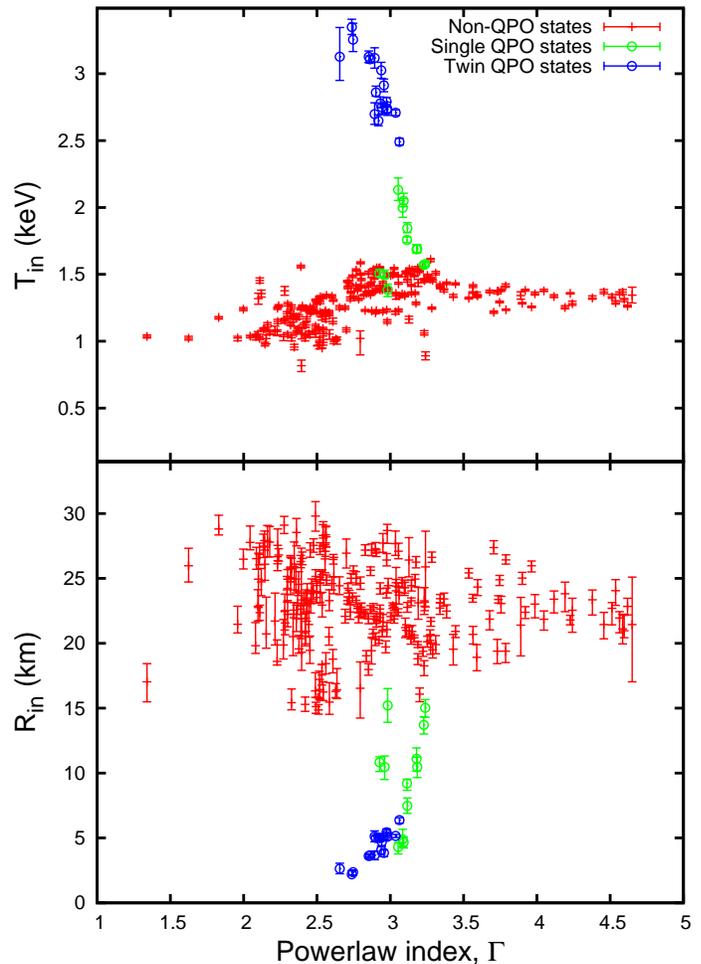}
\vspace{-6.0mm}
 \caption{The plot of inner disc temperature $T_{in}$ (top panel) and the inner disc radius $R_{in}$ (bottom panel) versus the powerlaw index $\Gamma$ for the QPO states. The red marks represent the non-QPO states. The green marks represent the single QPO states and the blue marks represent the twin QPO states. The three green marks forming a parallel branch correspond to flare 5A while the rest green marks correspond to flare 5B. The blue marks do not show any such demarcations.} 
 \label{fig_06}
\end{figure}

The two distinct QPO states, single and twin, suggest that the dynamics of the accretion geometry is extremely varied. The discontinuity of the $T_{in}-R_{in}$ behaviour indicate distinct accretion features hinting at different physical origin of the single QPO and the twin QPO states, despite the similarity in the frequency and other parameters of the Lorentzian functions of these QPOs. The distinctness of the two QPO states is also seen in the plot of $T_{in}$ as well as $R_{in}$ w.r.t. the \emph{power law} index $\Gamma$ (Fig. \ref{fig_06}). A significant feature is that in the bottom panel the three green marks embedded in the non-QPO states correspond to the flare 5A. This shows that the for the single QPO state, the $T_{in}$ forms a parallel track as a function of $\Gamma$ in different flaring episodes. The corresponding points on the top panel ($R_{in}$) also shows a parallel track for the green marks corresponding to the same flare 5A. The twin QPO states do not form parallel tracks despite being present during both the flares. Thus, it is clear that the evolution of emission from the accretion disc is more complex and depends on more than one physical origin, namely the accretion rate. The QPO states occur for a very narrow band of values of the \emph{powerlaw} index, $\Gamma$.

\subsection{The fluxes of the spectral model components}
\begin{figure*}
\includegraphics[width=0.78\textwidth, angle=-90]{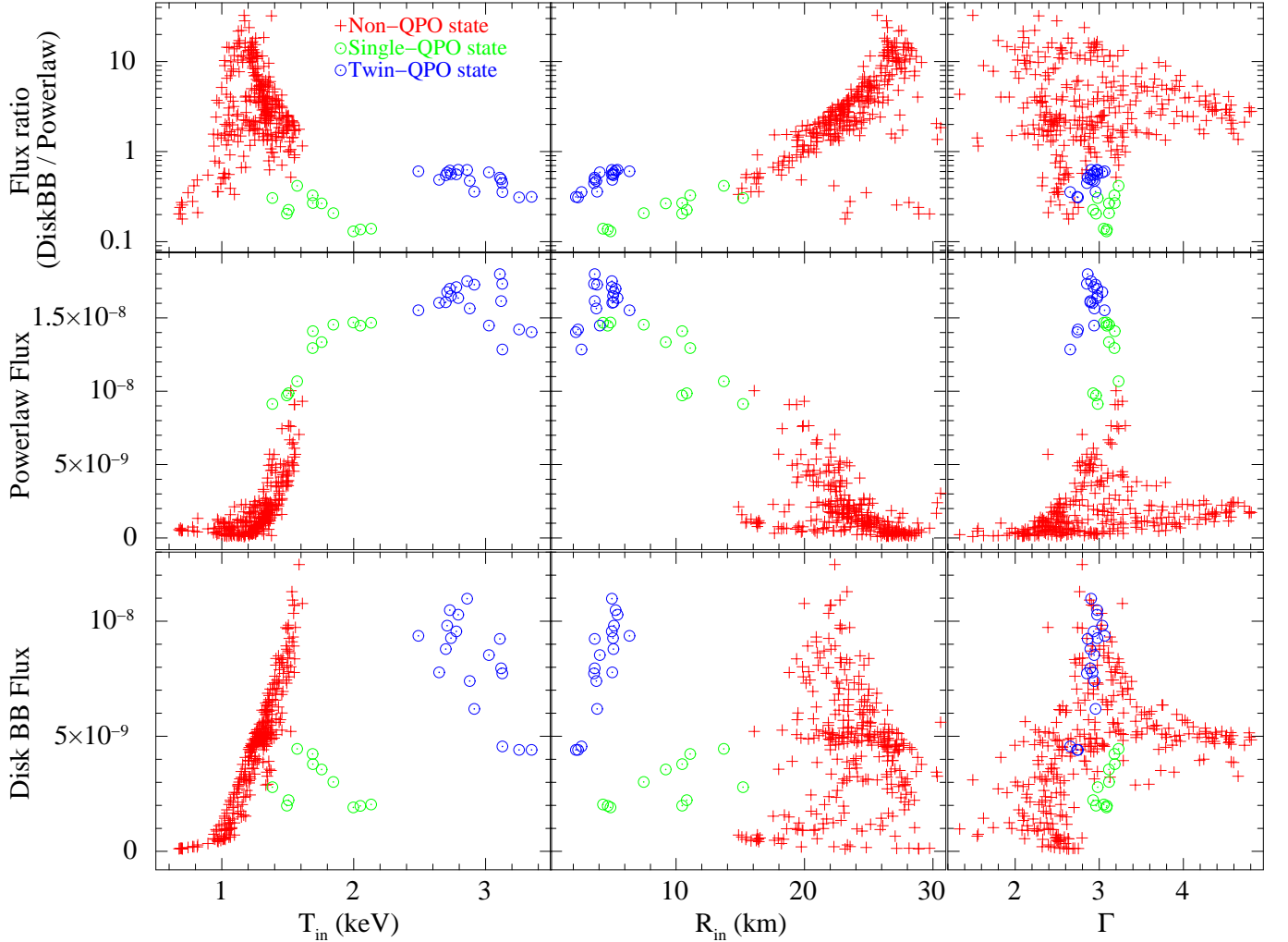}
\vspace{-2.5mm}
 \caption{The plot of the fluxes of the different model components of the energy spectra versus the inner disc temperature $T_{in}$ (left-hand column), the inner disc radius $R_{in}$ (middle column) and the \emph{power law} index $\Gamma$ (right-hand column). The multicoloured disc blackbody (\emph{diskbb}) flux is plotted on the bottom row, the \emph{power law} flux on the middle row and the ratio of the two fluxes on the top row. The red marks represent the non-QPO states. The green circled dots represent the single QPO states, the blue circled dots represent the twin QPO states. The three green circles forming a parallel branch (distinctly for the \emph{diskbb} flux and loosely for the total flux) correspond to flare 5A while the rest green circles correspond to flare 5B. } 
 \label{fig_07}
\end{figure*}

In Fig. \ref{fig_07} the scatter plot of fluxes of the model components, namely \emph{diskbb} (bottom row) and \emph{power law} (middle row) along with the ratio of the \emph{diskbb} flux to the \emph{power law} flux (top row) is plotted with the model parameters ($T_{in}$, $R_{in}$ and $\Gamma$). The red plus marks represent the non-QPO state, the green circled dots represent the single QPO states and the blue circled dots represent the twin QPO states.
It is evident that the \emph{diskbb} flux for the different QPO states form parallel branches for the single and twin QPO states, and furthermore, the single QPO data points also form parallel branches for the flare 5A and flare 5B (the three green circled dots forming a parallel branch w.r.t. the rest correspond to flare 5A, while the rest of the green circled dots correspond to flare 5B). The twin QPO state, in contrast, doesn't show any such demarcation despite being present during both the flares. This suggests that the emission from the thermal disc is very sensitive to the given local properties of flare in the case of the single QPO state in contrast to the twin QPO state. Furthermore, the \emph{power law} flux shows a continuous evolution as a function of the inner accretion disc parameters  as well as the \emph{power law} index (middle row for Fig. \ref{fig_07}). The \emph{power law} index, $\Gamma$, doesn't play any role in the emission from the disc blackbody (bottom right panel, Fig. \ref{fig_07}), but even here the different flux values for the single QPO state of flares 5A and 5B form distinct parallel tracks. 

It is very significant from the Fig. that the \emph{power law} flux is sensitive to the inner disk temperature, $T_{in}$, (Fig. \ref{fig_07} middle row) and not on the \emph{power law} index, $\Gamma$. Thus, the \emph{power law} flux is an excellent indicator of the presence of QPO state but not the \emph{power law} index $\Gamma$.

From Fig. \ref{fig_07} it is evident, from it's relation with the \emph{diskbb} flux as well as the \emph{power law} flux, that the deviation of the $T_{in}$ from its continuous evolution occurs during the QPO states. Furthermore, the QPO state is characterized by less disc blackbody (\emph{diskbb}) flux but more \emph{power law} flux. Hence, the contribution of the standard geometrically thin disc appears to be minimal in the case of the QPOs seen in the source (top row, Fig. \ref{fig_07}).
One interesting inference from Fig. \ref{fig_07} is that the extreme steep \emph{power law} observations do not correspond to the anomalous (QPO) state. 
Another important feature is that the disc blackbody flux evolves discontinuously as the source makes the transition from the non-QPO state to single QPO state to twin QPO state and back; whereas, the \emph{power law} flux undergoes continuous (and monotonic) change as the state transits across the three states defined by the presence and absence of QPOs.
Thus, it may be conjectured that saturated Comptonization and the appearance of slim disc might happen to be two opposing features, each competing to establish itself at the other's expense, and for this source the appearance of the QPO can be accepted as a signature of a geometrical evolution of the accretion disc.

\subsection{Correlation among the spectral parameters and fluxes}
To test for the correlation among the various model parameters (namely $T_{in}$, $R_{in}$, $\Gamma$) and the fluxes of the model components (\emph{diskbb} and the \emph{power law}), we have used the Spearman rank correlation (SRC) coefficient adapting the method prescribed by \citet{macklin82mnras}. For each correlation test between two variables a set of three variables is chosen and the corresponding D-parameter is derived, which gives the confidence level, in terms of standard deviations, that the derived correlation between the first two variables is not due to the influence of the third parameter. If the D-parameter has a value $>\mid\!\!1\!\!\mid$, then the possibility of the correlation of the two parameters can be ruled out as due to the third parameter. The important and significant cases from the results of this correlation test is given in table \ref{tab_02}. The tests were performed separately for the single and the twin QPO states (excluding MJD 52855). The obvious reason for the separation of the two states is that the two states give rise to parallel branches in the parameter space, hence the same value of one parameter will give rise to discontinuous multiple values of the other parameter, killing the opportunity of gleaning any useful information from the correlation test. Given the paucity of data points for the QPO states, the values obtained for the SRC, and especially the D-parameter, is suggestive at its best, but nevertheless it provides important insight into physical scenario of the accretion process.

\subsubsection{Single QPO state}
From table \ref{tab_02} it is evident that all the reported correlations are very strong when only individual flaring episode is considered (namely flare 5B; since flare 5A has only three data points its statistical parameters are not obtained, but a visual inspection of Fig. \ref{fig_06} and \ref{fig_07} suggests a strong correlated behaviour of the various parameters) for the single QPO state. Since in flare 5B the $T_{in}$ and $R_{in}$ are perfectly anti correlated, the SRC and the D-parameter involving them and any other parameter have the same value with different sign. For the individual flare 5B, the only pair of parameters that give inferior correlation as compared to the combined flaring episodes is $T_{in}$ and \emph{power law} flux. This is a statistical conundrum that requires more data, which is currently beyond the scope of the paper.

\begin{table}
\caption{The important and significant cases of the result of the Spearman partial correlation test among the model parameters and the fluxes of the model components, performed and presented separately for single and twin QPO states.}
\label{tab_02}
\medskip {Single QPO state (for flares 5A and 5B)(no. of data points: 11)\\}
\begin{tabular}{@{}lcccc}
Parameters & ($3^{rd}$ par) & SRC coeff. & Null prob. & D-par \\
\hline
$T_{in}$:$R_{in}$ & ($\Gamma$) &  -0.88 & $3\times10^{-4}$ & -4.83  \\
$R_{in}$:$\Gamma$ & ($T_{in}$) &   0.16 & $\ldots$ 		& $\ldots$ \\    
$T_{in}$:$\Gamma$ & ($R_{in}$) &   0.23 & $\ldots$       & $\ldots$ \\
$T_{in}$:\emph{diskbb} flux & ($R_{in}$/$\Gamma$) &   -0.33 & $\sim0.32$ & $>\mid\!1\!\mid$ \\
$T_{in}$:\emph{power law} flux & ($R_{in}$/$\Gamma$) & 0.95 & $\sim1\times10^{-5}$ & $>\mid\!1\!\mid$ \\
$R_{in}$:\emph{diskbb} flux & ($T_{in}$/$\Gamma$) &   0.64 &  $\sim0.03$  	& $>\mid\!1\!\mid$ \\
$R_{in}$:\emph{power law} flux & ($T_{in}$/$\Gamma$) & -0.82 & $\sim2\times10^{-3}$ & $>\mid\!1\!\mid$ \\
$\Gamma$:\emph{diskbb} flux & ($T_{in}$/$R_{in}$) &   0.64 &  $\sim0.03$  	& $>\mid\!1\!\mid$ \\
$\Gamma$:\emph{power law} flux & ($T_{in}$/$R_{in}$) & -0.82 & $\sim2\times10^{-3}$ & $>\mid\!1\!\mid$ \\

\hline \hline
\end{tabular}

\medskip {Single QPO state for only flare 5B (no. of data points: 8)\\}
\begin{tabular}{@{}lcccc}
Parameters & ($3^{rd}$ par) & SRC coeff. & Null prob. & D-par \\
\hline
$T_{in}$:$R_{in}$ & ($\Gamma$) &  -1.00 & 0.00 				&  $-\infty$    \\
$R_{in}$:$\Gamma$ & ($T_{in}$) &   0.93 & $8\times10^{-4}$	& $\ldots$ \\    
$T_{in}$:$\Gamma$ & ($R_{in}$) &  -0.93 & $8\times10^{-4}$	& $\ldots$ \\
$T_{in}$:\emph{diskbb} flux & ($\Gamma$) &   -0.90 & $2\time10^{-3}$ & -1.08 \\
$T_{in}$:\emph{power law} flux & ($\Gamma$) & 0.86 & $6\times10^{-3}$ & 1.06 \\
$R_{in}$:\emph{diskbb} flux & ($\Gamma$) &   0.90 & $2\time10^{-3}$ & 1.08 \\
$R_{in}$:\emph{power law} flux & ($\Gamma$) & -0.86 & $6\times10^{-3}$ & -1.06 \\
$\Gamma$:\emph{diskbb} flux & ($T_{in}$/$R_{in}$) &   0.88 & $3\times10^{-3}$  & 0.53 \\
$\Gamma$:\emph{power law} flux & ($T_{in}$/$R_{in}$) & -0.81 & 0.01 & -0.14 \\

\hline \hline
\end{tabular}
\medskip \textit{Since $T_{in}$ and $R_{in}$ are perfectly anti-correlated during the flare 5B, the SRC and the D-parameter involving them have the same value with different sign (same as the SRC value). Also if either of $T_{in}$ and $R_{in}$ is one of the main parameters to be tested and the other is the third parameter, then D-parameter is undefined.\\}
\medskip {\\ Twin QPO state (no. of data points: 18)\\}
\begin{tabular}{@{}lcccc}
Parameters & ($3^{rd}$ par) & SRC coeff. & Null prob. & D-par \\
\hline \hline
$T_{in}$:$R_{in}$ & ($\Gamma$) &  -0.91 & $1\times10^{-7}$ &  -4.12  \\
$R_{in}$:$\Gamma$ & ($T_{in}$) &   0.83 & $2\times10^{-5}$ &  2.41 \\    
$T_{in}$:$\Gamma$ & ($R_{in}$) &  -0.74 & $5\times10^{-4}$ &  0.35 \\ 
$T_{in}$:\emph{diskbb} flux & ($R_{in}$/$\Gamma$) &  $3\times10^{-3}$5 & $\ldots$ & $>\mid\!1\!\mid$ \\
$T_{in}$:\emph{power law} flux & ($R_{in}$/$\Gamma$) & 0.20 & $\ldots$ & $\ldots$ \\
$R_{in}$:\emph{diskbb} flux & ($T_{in}$/$\Gamma$) & 0.80 & $7\times10^{-5}$ & $>\mid\!1\!\mid$ \\
$R_{in}$:\emph{power law} flux & ($T_{in}$/$\Gamma$) & 0.22 & $\ldots$ & $\ldots$ \\
$\Gamma$:\emph{diskbb} flux & ($R_{in}$) & 0.62 & 0.36 & 0.30 \\
$\Gamma$:\emph{diskbb} flux & ($T_{in}$) & 0.62 & $6\times10^{-3}$ & 1.03 \\
$\Gamma$:\emph{power law} flux & ($R_{in}$/$T_{in}$) & 0.23 & $\ldots$ & $\ldots$ \\
\hline \hline
\end{tabular}
\end{table} 

While the anti correlation of $T_{in}$ and $R_{in}$ is obvious from the spectra (Fig. \ref{fig_03} and \ref{fig_05}), the anti correlation of the $T_{in}$ and the \emph{diskbb} flux (or conversely, the correlation of the $R_{in}$ and the \emph{diskbb} flux) is a bit anti-intuitive, as it is expected that with the accretion disc extending further inwards the flux from the thermal disc component will increase in proportion, but just the opposite happens. In contrast, as the disc supposedly extends inwards, the \emph{power law} flux increases (see values of $T_{in}$:\emph{power law} flux and $R_{in}$:\emph{power law} flux in table \ref{tab_02}, also see Fig. \ref{fig_07}). Furthermore, the \emph{power law} flux has a very strong anti correlation with $\Gamma$, while the \emph{diskbb} flux is very strongly correlated with $\Gamma$. This coincides with the fact that the \emph{power law} index $\Gamma$ is very strongly anti correlated with the inner disc temperature $T_{in}$ and positively correlated with the inner disc radius $R_{in}$, i.e. as the disc extends inwards the \emph{power law} hardens and its emission increases. Conversely, the \emph{diskbb} flux increases as the inner disc extends outwards and the \emph{power law} index, $\Gamma$, steepens. Important point to be remembered is that the QPO states exist within a very narrow band of the values of $\Gamma$. Hence, in the QPO states, the \emph{power law} index, $\Gamma$, is very sensitive to the accretion disc related properties and behaves in a very counter intuitive manner which is different from the normal behaviour seen in the non-QPO states (Fig. \ref{fig_06} and \ref{fig_07} bottom right and middle right panels).

The D-parameter suggests that the spectral parameters (viz. $T_{in}$, $R_{in}$, $\Gamma$) are all independent, while the correlation of the \emph{power law} flux with the \emph{power law} index, $\Gamma$, is most likely caused by the inner disc parameters. This suggest that the extent of the disc is more important a player compared to the presence of the Compton cloud, during the single QPO state. Additionally some significant change in the physical and/or geometrical state of the disc is being manifested in the inner disc parameters and their correlation with the model component fluxes.

\subsubsection{Twin QPO state}
During the twin QPO state, the one parameter that is completely different in values from the single QPO state is the inner disc temperature $T_{in}$ (Fig. \ref{fig_07}). 
In general, the data points in the scatter diagram of parameter values and the model component fluxes are more clustered than being correlated for the twin QPO state (table \ref{tab_02} and Fig. \ref{fig_07}), especially for the relation of flux values with $T_{in}$.  Also, the clustering in different flaring episodes do not give rise to discernible branches in the scatter plots, nor are the correlation parameters significantly altered if the analysis is done separately for the individual flaring episodes. But, since the clustering of the data points happen in distinct regions, it is quite possible that the QPO states have different physical origins. Yet, the general trend suggest somewhat similar evolution of the disc and the Compton cloud and their emission as to the single QPO state.

The model parameters  $T_{in}$, $R_{in}$ and $\Gamma$ show equally strong and independent correlation result as in the case of single QPO state. The significant difference of the SRC values from  that of the single QPO state is that the \emph{power law} flux is not correlated to any of the model parameters, while the \emph{diskbb} flux has weaker correlation compared to the single QPO state. The D-parameter suggest that the $\Gamma$:\emph{diskBB} flux correlation is due to the third parameter $R_{in}$ (and independent of $T_{in}$). 

Within the twin QPO state, the \emph{power law} flux reaches a saturation w.r.t. $T_{in}$, while the \emph{diskbb} flux falls with increasing $T_{in}$ (Fig. \ref{fig_07}). But, in comparison with the single QPO state, the flux of both the components (\emph{diskbb} and \emph{powerlaw}) is higher for the twin QPO state. From the top row of Fig. \ref{fig_07} it is evident that during the twin QPO state the \emph{diskbb} flux increases discontinuously. Hence, the twin QPO state, in a sense, depicts a more extreme version of the evolution of the accretion disc and the Compton cloud.

\section{Frequency dependence of spectral parameters}

\begin{figure}
\includegraphics[width=0.525\textwidth]{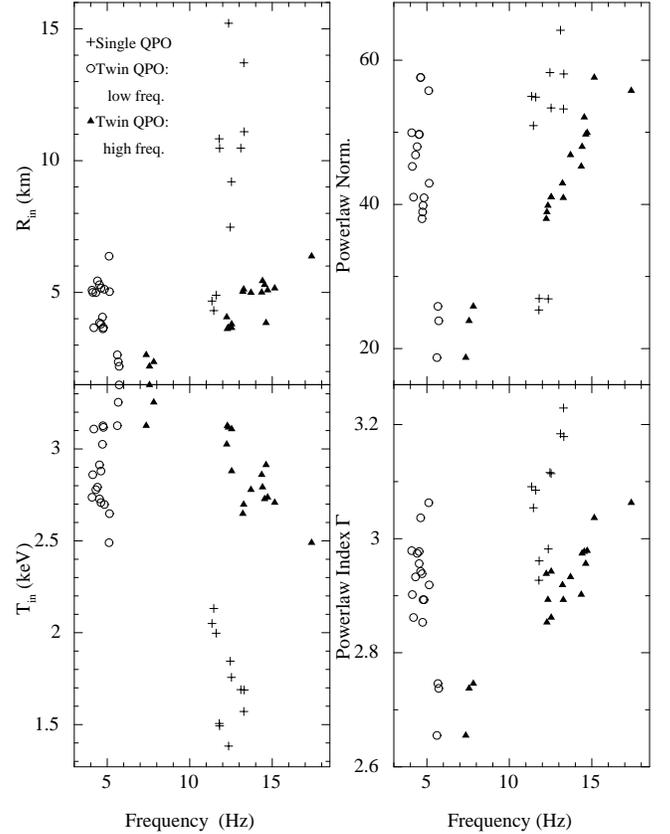}
\vspace{-11.0mm}
 \caption{The various parameters of the model components that define the energy spectra for the QPO states versus the frequency of the QPO. Plus marks represent the single QPO state; filled triangles represent the higher frequency of twin QPO state; circles represent the lower frequency of the twin QPO state. } 
 \label{fig_08}
\end{figure}

\begin{figure}
\includegraphics[width=0.69\textwidth, angle=-90]{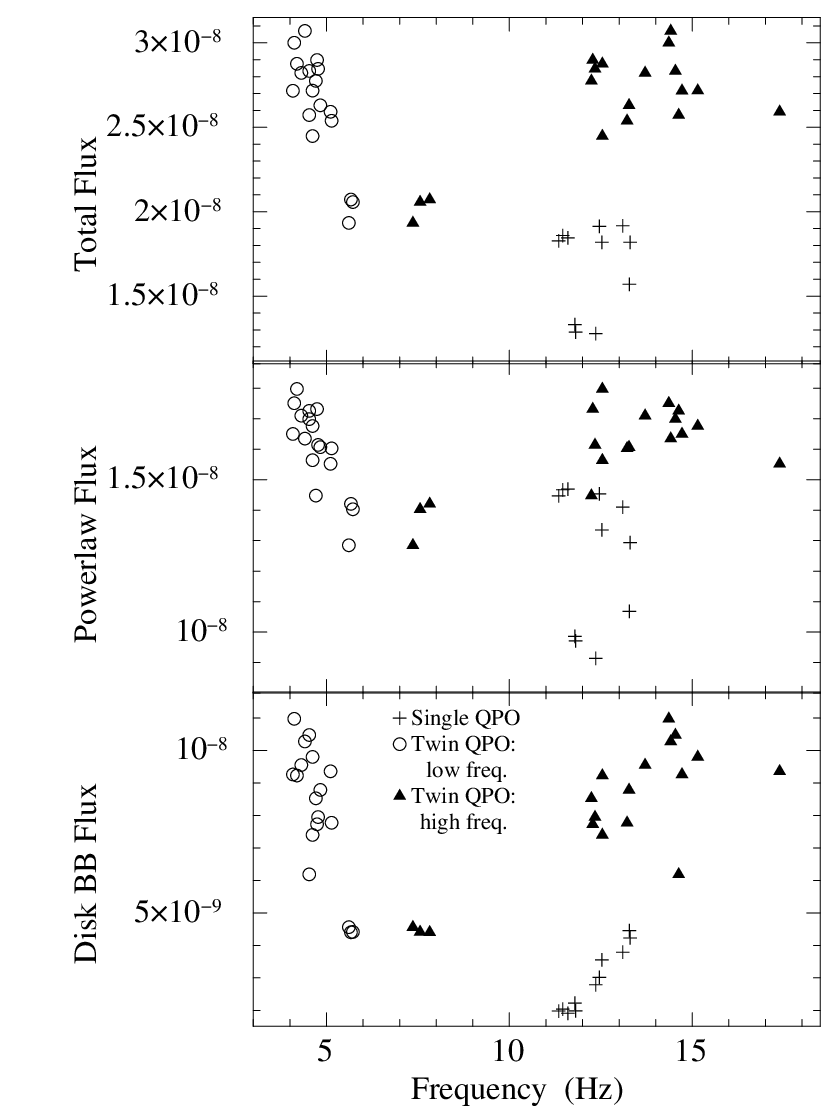}
\vspace{-2.5mm}
 \caption{The flux values of the model components that define the energy spectra for the QPO states versus the frequency of the QPO. Plus marks represent the single QPO state; filled triangles represent the higher frequency of twin QPO state; circles represent the lower frequency of the twin QPO state. } 
 \label{fig_09}
\end{figure}

The QPO frequency dependence of the various parameters and flux values are essential for further insight to the accretion system. 
In Fig. \ref{fig_08} the scatter diagram of the various spectral parameters ($T_{in}$, $R_{in}$, $\Gamma$, Powerlaw norm) are plotted with the centroid frequencies of the various QPOs. In Fig. \ref{fig_09} the scatter diagram of the flux values of the spectral model components are plotted with the frequencies. In concordance with the results mentioned in the previous section, the spectral parameters as well as the flux values cluster in different regions for the different frequencies of the QPOs. The SRC coefficient is computed for the correlation between the centroid frequency of the various QPOs and the spectral parameters as well as the flux values. The SRC is computed using any dummy $3^{rd}$ parameter as in this case the computation of the D-parameter is not helpful. The SRC coefficients are tabulated in table \ref{tab_03}.

\begin{table*}
\caption{The Spearman partial correlation (SRC) coefficient of the correlation between the QPO frequencies and the model parameters and the model component fluxes. The value of D-parameter given in brackets with the $3^{rd}$ parameter when it is less than 1.}
\label{tab_03}
\begin{tabular}{@{}llcclc}
\hline \hline
Frequency type & $T_{in}$ & $R_{in}$ & $\Gamma$ & \emph{diskbb} flux & \emph{power law} flux \\
\hline
Frequency: single QPO state & -0.44 & 0.68 & 0.69  & 0.93 & -0.34  \\
(both flares 5A and 5B) & & & & & \\
Frequency: single QPO state & -0.95 & 0.95 & 0.83 & 0.90 & -0.76  \\
(flare 5B only) & & & & & \\
Higher frequency: twin QPO state & -0.80 (-0.55/$R_{in}$) & 0.85 & 0.88 & 0.70 (0.32/$R_{in}$) & 0.47 \\
Lower frequency: twin QPO state & 0.27 & -0.40 & -0.48 & -0.69 & -0.79 \\
\hline \hline
\end{tabular}
\medskip \\ Notes. The SRC coefficients are obtained using any dummy $3^{rd}$ parameter when D-parameter is greater than 1.

\end{table*} 

\subsection{Single QPO state}
The correlation of the frequency with the various parameters in this state is much stronger when the individual flare 5B is considered (table \ref{tab_03}). The frequency scales very strongly with the $R_{in}$ and inversely with the $T_{in}$. Therefore, in the course of the evolution of the disc, the QPO frequency decreases as the disc extends inwards and subsequently the inner disc temperature $T_{in}$ rises. The correlation of the frequency with \emph{power law} index, $\Gamma$, is very strong, but weaker in comparison to the correlation with the inner disk parameters.
The frequency is also very strongly correlated to the \emph{diskbb} flux and strongly anti correlated to the \emph{power law} flux.

\subsection{Twin QPO state}
The twin QPO state show very distinct behavioural features of the two frequencies. The two frequencies show a weak anti correlation (SRC=--0.56) between themselves. The behaviour of the higher frequency (7-18 Hz) is somewhat similar to the single QPO frequency, with the SRC coefficient being less significant as the scatter of the parameters is more in the twin QPO state. Nonetheless, the frequency scales with the inner disc radius $R_{in}$ and inversely so with the inner disc temperature $T_{in}$. But, the correlation of the frequency is highest with the \emph{power law} index, $\Gamma$. This is in contrast to the single QPO state. This frequency is also correlated with the \emph{diskbb} flux, but with the \emph{power law} flux the SRC coefficient value depicts a lack of correlation, hinting at a high scatter of this quantity w.r.t the frequency (but it is definitely not anti correlated, as was the case for the single QPO state). The D-parameter suggest this frequency is correlated to the $T_{in}$ and the \emph{diskbb} flux by the parameter $R_{in}$.

The lower frequency (4-6 Hz) does not correlate with the model parameters (table \ref{tab_03}), although it forms a separate tight cluster in the scatter diagram (Fig. \ref{fig_08}). Intriguingly, the flux of the model components, however, are strongly anti-correlated to this frequency.


\section{Evolution of the accretion states}
Given the results obtained in the previous sections, it is interesting to note the evolution of the structure of the accretion disc. Starting from the quiescent state, the steady build up of both the flares 5A and 5B (Fig. \ref{fig_02}) result in the standard Shakura--Sunyaev disc extending slowly inwards, along with the presence of the Comptonizing cloud (the \emph{power law} component). The flares 5C, 5D, and 5E also behave in the same manner. 
But for flares 5A and 5B the rise of the flare does not stop with the luminosity peaking at  $\sim 0.4$ Crab flux (like it does for flares 5C, 5D, and 5E), but continues to increase. This  most likely leads to the QPO states when the applicability of the geometrically thin disc model becomes entirely infeasible. 

The temporal evolution may consist of the following stages. 1) The Shakura--Sunyaev disc extends inwards beyond a certain threshold with the fall in $R_{in}$ and rise in $T_{in}$. The thin disc emission reduces and the \emph{power law} emission (which can be called a non-thin-disc emission) rises. The onset of single QPO state happens and as the $R_{in}$ falls the value of the single QPO frequency also falls while the $T_{in}$ rises along with the hardening of the \emph{power law} component, and the \emph{diskbb} flux correspondingly falls while there is a rise in the \emph{power law} flux. 2) The process may undergo a turn around and start evolving in the opposite direction, with a rise in $R_{in}$ and a corresponding fall in $T_{in}$, in addition to all the other correlated changes in the parameters of the model and the QPO, resulting in the corresponding increase in the \emph{diskbb} flux and decrease of the \emph{power law} flux.
3) With further change in the accretion and/or disc parameters in the single QPO state, the onset of the twin QPO state occurs. The spectral parameters behave similarly, albeit at a definite higher value of $T_{in}$ and lower value of $R_{in}$, suggesting that the accretion disc extends further inwards. 
The \emph{power law} index $\Gamma$ remains within the narrow limit analogous to the single QPO state. But the \emph{power law} flux does not depend on the spectral parameters, while the  \emph{diskbb} flux now anti-correlates with $\Gamma$, the likely cause being the inner disc radius $R_{in}$. Here the higher frequency of the two QPOs behaves somewhat similarly to the frequency of the single QPO state, i.e. it scales with the $R_{in}$ and inversely with $T_{in}$, but the correlation of the frequency is highest with $\Gamma$.  It is the lower frequency that causes the system to behave in an inexplicable manner. The value of the frequency is anti-correlated with both the thin disc flux (\emph{diskbb}) as well as the \emph{power law} flux. 4) The system may fluctuate within the twin QPO state with corresponding changes in the parameters and fluxes, or it might transit down to the single QPO state. From the single QPO state it might again make a transition to the twin QPO state or it may make the transition back to the canonical state as the flare dies down.

Figs \ref{fig_02} and \ref{fig_03} (bottom panel) show very fast fluctuation of the X-ray emission during the peak phase of the flares 5A and 5B, and this feature is reflected in the spaced out data points of the QPO states in the $R_{in}$--$T_{in}$ plane (Fig. \ref{fig_05}). 
This fluctuation of the flux (Figs \ref{fig_02} \& \ref{fig_03}) occur independent of any change in the mass accretion rate as the slope of the log-log diagram of $R_{in}$--$T_{in}$ (Fig. \ref{fig_05}) remains constant \citep{watarai00pasj,watarai01apj}, provided a transition from the geometrically thin to geometrically slim disc has taken place. In this case the fluctuation of luminosity may suggest some instability in the accretion disc causing the drastic fluctuation of the $R_{in}$ and $T_{in}$ along with the onset of the QPOs, or else a fast physical mechanism is removing the mass from the inner disc close to the event horizon, resulting in the truncation of the disc and lowering of luminosity (Fig. \ref{fig_07} bottom row). Eventhough the geometrically thin disc model is incapable of properly parametrizing the comparatively thicker slim disc parameters, but the total luminosity measured from the light curves (3-30 keV for PCA and 1.5--12 keV for \textit{ASM}) is independent of the spectral model, suggesting that the fluctuation of the disc is a real phenomenon.

This evolution may perhaps indicate that the geometrically thin disc component is actually incapable of giving proper explanation of the spectral evolution, and the \emph{power law} component of the total flux may contain the emission actually originating from the thicker part of the slim disc. This may also explain the rise of the \emph{power law} flux with the rise in $T_{in}$ and fall in $R_{in}$ as well as the \emph{diskbb} flux. The onset of the geometrically slim disc may also explain the residue around 8--10 keV that is seen in the QPO states (see Section 2). 

\begin{figure}
\includegraphics[width=0.5\textwidth]{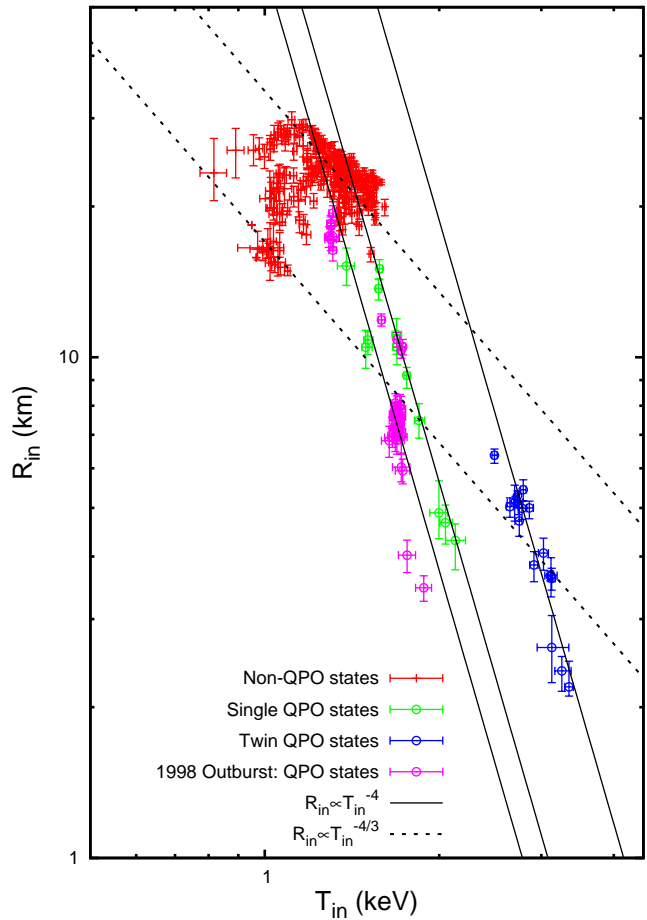}
\vspace{-7.5mm}
 \caption{The plot of inner disc radius $R_{in}$ versus inner disc temperature $T_{in}$ for all the QPO observations of the 1998 outburst (peak phase)along with the 2002-2004 outburst (Fig. \ref{fig_05}). The values of $R_{in}$ and $T_{in}$ are obtained from the spectral parameters reported by \citet{trudolyubov01mnras} (see the text for details). The red marks represent the QPO states of the 2002-2004 outburst; the green and blue marks represent the QPO sates; the purple marks represent the QPO observations of the 1998 outburst. The non-QPO observations obey the relationship $R_{in} \propto T_{in}^{-4/3}$ (geometrically thin optically thick Shakura--Sunyaev disc). The QPO observations obey the relation $R_{in} \propto T_{in}^{-4}$. } 
 \label{fig_10}
\end{figure}


\subsection{Comparison with 1998 Outburst}
The 1998 outburst of 4U 1630--47 (outburst 2 as per \citealt{abe05pasj}) is well reported in the literature \citep{hjellming99apj,dieters00apj,tomsick00apj}. The duration of this outburst is around 100 d with one major flare with peak flux $\sim 0.4$ Crab lasting for about 60 d and a very small follow up flare with peak flux $\sim 0.1$ Crab flux, that lasted for about 40 d. Using the \textit{RXTE}/PCA data \citet{trudolyubov01mnras} reported the presence of QPOs during the peak phase of this outburst which lasted for about 30 d. They also reported the best-fitting energy spectral parameters of the same \citep[see table 2 of][]{trudolyubov01mnras}. The model used was essentially very similar to our model used here, the only difference being the addition of a \emph{cut-off \emph{power law}} to fit the 19-100 keV data obtained from \textit{RXTE}/HEXTE; the \textit{RXTE/PCA} was used to obtain the spectra in the range of 2-20 keV. From their best-fitting parameters we have calculated the inner disc radius, $R_{in}$ for the QPO-observations of the 1998 outburst, assuming the source to be at distance of 10 kpc and the binary system inclined at an angle of $60^o$. The resultant $T_{in} - R_{in}$ plot for the 1998 outburst is plotted in Fig. \ref{fig_10} where the data for the 2002-2004 outburst (erstwhile Fig. \ref{fig_05}) is also plotted for comparison. The three parallel lines representing the QPO states stand out in Fig. \ref{fig_10},  which suggest that the behaviour has extremely close resemblance (albeit qualitative) with the behaviour expected in the presence of a slim disc \citep{watarai00pasj, watarai01apj}. During this outburst the presence of harmonically unrelated multiple QPOs in individual observations is reported \citep[see table 3 of][]{trudolyubov01mnras}. Hence, the outburst 2 might be very similar to the outburst 5 (flares 5A and 5B) at a lower mass accretion rate.

Interestingly, the follow up flares (5C, 5D, 5E, and also perhaps 2B) do not exhibit any quasi-periodic behaviour, while their energy spectral parameters also conform to the geometrically thin approximation. Perhaps outbursts with relatively strong mass accretion rate can only give rise to the quasi-periodic behaviour, which appears to be a strong indicator of the formation of slim disc.

\begin{figure}
\includegraphics[width=0.35\textwidth, angle=-90]{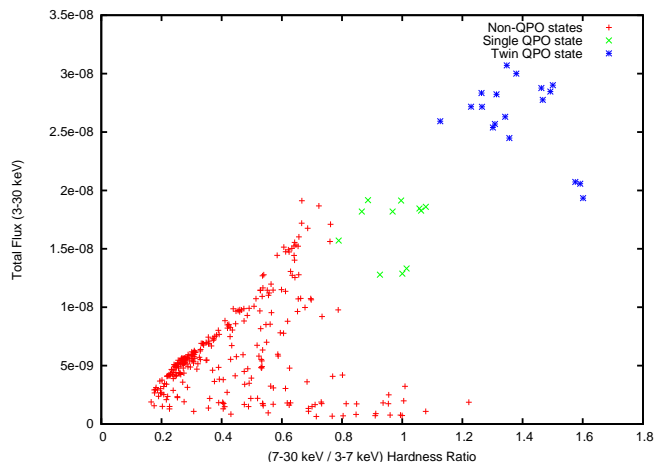}
\vspace{-3.0mm}
 \caption{The intensity-hardness ratio plot for the source 4U 1630--47 during the outburst 5. The red plus marks represent the non-QPO state, the green crosses represent the single QPO state and the blue crosses represent the twin QPO state.
} 
 \label{fig_11}
\end{figure}


\section{Discussion}
The widely diverse nature of the emission from the Galactic black hole binaries require various classifications schemes to make sense of their radiation mechanism and the underlying physical mechanism \citep[and the references therein]{done07aarv}. The component based classification scheme classifies all the `possible' spectral states of the black hole systems into five states \citep{mcclintock06}. On the other hand the model independent classification scheme of \citet{homan05apss} pays more stress on the transition between the states (Fig. \ref{fig_11}) and the related radio emission from the outflow from the  system. The component based classification scheme \citep{mcclintock06} uses a multicoloured optically thick geometrically thin accretion disc model (best parametrized by the simplistic \emph{diskbb} model) with a combination of a \emph{power law} and/or \emph{cutoff powerlaw}. 
Hence the basic premise of the whole classification scheme is based on a geometrically thin disc. As a fallout, all the popular complex models used to fit the X-ray emission spectra are implicitly based on the assumption of the accretion disc being geometrically thin \citep[see for eg.][]{kubota98pasj,gierlinski99mnras,zimmerman05apj,gou11apj}. Although the classification of \citet{homan05apss} is based on the colour--intensity evolution during the state transition of the sources and hence independent of the specific physical model used to describe the spectra, the physical interpretation of these transitions have an implied assumption of geometrically thin disc \citep{belloni05aipc, belloni11basi}. Comparatively, the attempt to fit observational data to the slim disc model has been sparse and few \citep[namely][]{watarai01apj}.

The classification of the low-frequency QPOs into three main types \citep{wijnands99apjl,homan01apjs} for the black hole binary system is quite established \citep{casella05apj}. This classification is helpful in providing greater depth to the understanding of the energy spectral component based classifications described above. This A-B-C classification scheme normally requires the presence of one fundamental frequency along with its harmonic (or at times the sub harmonic) \citep{casella04aa}. The presence of multiple QPOs with frequencies not harmonically related in the source 4U 1630--47 adds to the complexity of classification of these states. But, the observation of simultaneous non-harmonically related QPOs is not unique to the source \citep{motta12arXiv1209.0327}. A complete understanding of all the QPO observations covering the various types of outbursts of the source continuously monitored by the \textit{RXTE} is needed to properly classify the QPOs observed in this source.

\citet{rao00aa} do suggest that the low frequency QPO may occur due to the formation of shock in the transition region of the geometrically thin but optically thick accretion flow and the low angular momentum optically thin accretion flow. Thus the possibility of high inner disc temperature occurring due to electron scattering can not be ruled out at the moment, with the seed photon of the Compton scattering originating at the optically thick geometrically thin disc and the Compton up scatter occurring in the optically thin cloud. Nevertheless, the failure of the Comptonzation models to fit the energy spectral data with physically meaningful parameters suggests that the $T_{in} -R_{in}$ behavioural pattern in Fig. \ref{fig_05} and \ref{fig_10} is qualitatively correct.
The similarity between the $T_{in}$--$R_{in}$ relationship and the model provided by \citet{watarai00pasj} provide the formation of slim disc as a very attractive physical option hitherto not explored. Detailed physical mechanism of the thick accretion disc need to be modelled to explore the possible physical mechanisms that may give rise to the QPO phenomena \citep[namely diskosiesmology;][]{wagoner12apjl}. The possibility of X-ray emissions from the jet during the anomalous state \citep{trigo13nat} requires the explanation of the possibility of the QPO phenomenon occurring in the outflow.

The observed coincidence of the appearance of the QPOs and the onset of slim disc, with or without the outflow emitting in X-rays, opens up many new possibilities in the modelling of the accretion disc for this source in particular and other black hole binary systems in general. The need of the hour is to develop a generalized multicoloured accretion disc model that would enable a smooth transition from the geometrically thin to the geometrically 'thicker', i.e. slim, disc, along with the emission from the out flowing jets. 


\section{Conclusion}
In this paper we report the onset of an anomalous accretion state during at least two of the many outbursts exhibited by the black hole binary system 4U 1630--47. This anomalous state is very closely associated with the appearance of low frequency QPOs. The source of the QPO during the unusual state needs to be explored theoretically. One possibility is the formation of shock in the transition region of the thermalized multicoloured geometrically thin and optically thick accretion disc and the Comptonizing plasma cloud which is geometrically thick and optically thin. The other option is the change in the physical and geometrical state of the accretion disc and the thin disc transforming into a thick disc with increase in accretion rate. 

For this source the outburst may consist of more than one flaring events. During each flaring event, the mass accretion is perhaps the most legitimate cause of the peak value of the flux measured from the flare. From the two outbursts covered (1998 and 2002-2004), it appears that the onset of the slim disc, if it appears, happens if the flux approaches a value of $\sim 0.4$ Crab and beyond. Otherwise the geometry of the disc remains thin and the non-thermal emission is best explained to originate from the optically thin Comptonizing cloud, and/or from the shock region at the boundary of the accretion disc and the Compton cloud. In such a scenario the formation of the slim disc and the existence of the Compton cloud might be competing events: but the last word in this can  not be claimed from the given data reported here. The $T_{in}$--$R_{in}$ behavioural patterns qualitatively suggest that for each flare the mass accretion perhaps does not change much but the slim disc phase sees a lot of fluctuation in terms of $R_{in}$ as well as $T_{in}$. This change in the inner disc temperature and the inner disc radius may be due to instability on the disc or the disappearance of matter from the inner part of the disc.


\section*{Acknowledgements} This research has made use of data obtained through
the HEASARC Online Service, provided by the NASA/GSFC, in support of NASA High
Energy Astrophysics Programs. MC acknowledges the many discussion with suggestions provided by Proffessor A R Rao, TIFR, Mumbai, India. This research has made use of the General High-energy Aperiodic Timing Software (GHATS) package developed by T. M. Belloni at INAF -- Osservatorio Astronomico di Brera.


\label{lastpage}


\begin{thebibliography}{}
\bibitem[\protect\citeauthoryear{Abe et al.}{2005}]{abe05pasj} 
Abe Y., Fukazawa Y., Kubota A., Kasama D., Makishima K., 2005, PASJ, 57, 
629 

\bibitem[\protect\citeauthoryear{Abramowicz et al.}{1988}]{abramowicz88apj} Abramowicz M.~A., Czerny B., Lasota J.~P., Szuszkiewicz E., 1988, ApJ, 332, 646 

\bibitem[\protect\citeauthoryear{Arnaud}{1996}]{arnaud96aspc} Arnaud 
K.~A., 1996, in Jacoby G. H., Barnes J., eds, ASP Cnf. Ser., Vol. 101, Astronomical Data
Analysis Software and Systems V. Astron. Soc. Pac., San Francisco, p.17 

\bibitem[\protect\citeauthoryear{Augusteijn, Kuulkers, \& van Kerkwijk}{2001}]{augusteijn01aa} Augusteijn T., Kuulkers E., van Kerkwijk M.~H., 2001, A\&A, 375, 447 

\bibitem[\protect\citeauthoryear{Belloni}{2005}]{belloni05aipc} Belloni T., 2005, n Burderi L., Antonelli L. A., D’Antona F., Di Salvo T., Israel G. L., Piersanti L.,
Tornamb`e A., Oscar Straniero O., eds, AIP Cnf. Proc., Vol. 797, Interacting Binaries: Accretion, Evolution, and Outcomes. Am. Inst. Phys., New York, p.197 

\bibitem[\protect\citeauthoryear{Belloni, Motta, \& Mu{\~n}oz-Darias}{2011}]{belloni11basi} Belloni T.~M., Motta S.~E., Mu{\~n}oz-Darias T., 2011, Bull. Astron. Soc. India, 39, 409 

\bibitem[\protect\citeauthoryear{Belloni, Psaltis, \& van der Klis}{2002}]{belloni02apj} Belloni T., Psaltis D., van der Klis M., 2002, ApJ, 572, 392 

\bibitem[\protect\citeauthoryear{Bloser et al.}{1996}]{bloser96aas} Bloser P.~F., Barret D., Grindlay J.~E., Zhang S.~N., Harmon B.~A., Fishman G.~J., Paciesas W.~S., 1996, A\&AS, 120, 191 

\bibitem[\protect\citeauthoryear{Bradt, Rothschild, \& Swank}{1993}]{bradt93aas} Bradt H.~V., Rothschild R.~E., Swank J.~H., 1993, A\&AS, 97, 355 

\bibitem[\protect\citeauthoryear{Casella, Belloni, \& Stella}{2005}]{casella05apj} Casella P., Belloni T., Stella L., 2005, ApJ, 629, 403 

\bibitem[\protect\citeauthoryear{Casella et al.}{2004}]{casella04aa} Casella P., Belloni T., Homan J., Stella L., 2004, A\&A, 426, 587 

\bibitem[\protect\citeauthoryear{Chen, Shrader, \& Livio}{1997}]{chen97apj} Chen W., Shrader C.~R., Livio M., 1997, ApJ, 491, 312 

\bibitem[\protect\citeauthoryear{Dieters et al.}{2000}]{dieters00apj} Dieters S.~W., et al., 2000, ApJ, 538, 307 

\bibitem[\protect\citeauthoryear{Done, Gierli{\'n}ski, \& Kubota}{2007}]{done07aarv} Done C., Gierli{\'n}ski M., Kubota A., 2007, A\&AR, 15, 1 

\bibitem[\protect\citeauthoryear{Gierli{\'n}ski et al.}{1999}]{gierlinski99mnras} Gierli{\'n}ski M., Zdziarski A.~A., 
Poutanen J., Coppi P.~S., Ebisawa K., Johnson W.~N., 1999, MNRAS, 309, 496 

\bibitem[\protect\citeauthoryear{Gou et al.}{2011}]{gou11apj} Gou L., et al., 2011, ApJ, 742, 85 

\bibitem[\protect\citeauthoryear{Hjellming et al.}{1999}]{hjellming99apj} Hjellming R.~M., et al., 1999, ApJ, 514, 383 

\bibitem[\protect\citeauthoryear{Homan \& Belloni}{2005}]{homan05apss} Homan J., Belloni T., 2005, Ap\&SS, 300, 107 

\bibitem[\protect\citeauthoryear{Homan et al.}{2001}]{homan01apjs} 
Homan J., Wijnands R., van der Klis M., Belloni T., van Paradijs J., 
Klein-Wolt M., Fender R., M{\'e}ndez M., 2001, ApJS, 132, 377 

\bibitem[\protect\citeauthoryear{Houck \& Denicola}{2000}]{houck00aspc} Houck J.~C., Denicola L.~A., 2000, in Manset N., Veillet C., Crabtree D., eds, ASP Conf. Ser., Vol. 216, Astronomical Data Analysis Software and Systems IX. Astron. Soc. Pac., San Francisco, p.591 

\bibitem[\protect\citeauthoryear{Houck}{2002}]{houck02} Houck J.~C., 2002, in Branduardi-Raymont G., ed., Proceedings of the International Workshop, High Resolution X-ray Spectroscopy with XMM-Newton and Chandra. Mullard Space Science Laboratory of University College London, Holmbury St Mary, Dorking, Surrey, UK, p. 24

\bibitem[\protect\citeauthoryear{Jones et al.}{1976}]{jones76apj} 
Jones C., Forman W., Tananbaum H., Turner M.~J.~L., 1976, ApJ, 210, L9 

\bibitem[\protect\citeauthoryear{Kalemci et al.}{2004}]{kalemci04apj} Kalemci E., Tomsick J.~A., Rothschild 
R.~E., Pottschmidt K., Kaaret P., 2004, ApJ, 603, 231 

\bibitem[\protect\citeauthoryear{Kubota \& Makishima}{2004}]{kubota04apj} Kubota A., Makishima K., 2004, ApJ, 601, 428 

\bibitem[\protect\citeauthoryear{Kubota et al.}{1999}]{kubota99an} Kubota A., Marshall F., Makishima K., Dotani T., Ueda Y., Negoro H., 1999, 
Astron. Nachr., 320, 353 

\bibitem[\protect\citeauthoryear{Kubota et al.}{1998}]{kubota98pasj} Kubota A., Tanaka Y., Makishima K., Ueda Y., Dotani T., Inoue H., Yamaoka 
K., 1998, PASJ, 50, 667 

\bibitem[\protect\citeauthoryear{Kuulkers, van der Klis, \& Parmar}{1997b}]{kuulkers97apjl} Kuulkers E., van der Klis M., Parmar A.~N., 1997b, ApJ, 474, L47 

\bibitem[\protect\citeauthoryear{Kuulkers et al.}{1998}]{kuulkers98apj} Kuulkers E., Wijnands R., Belloni T., Mendez M., van der Klis M., van Paradijs J., 1998, ApJ, 494, 753 

\bibitem[\protect\citeauthoryear{Kuulkers et al.}{1997a}]{kuulkers97mnras} Kuulkers E., Parmar A.~N., Kitamoto S., Cominsky L.~R., Sood R.~K., 1997a, MNRAS, 291, 81 

\bibitem[\protect\citeauthoryear{Levine et al.}{1996}]{levine96apjl} 
Levine A.~M., Bradt H., Cui W., Jernigan J.~G., Morgan E.~H., Remillard R., 
Shirey R.~E., Smith D.~A., 1996, ApJ, 469, L33 

\bibitem[\protect\citeauthoryear{Makishima et al.}{1986}]{makishima86apj} Makishima K., Maejima Y., Mitsuda K., 
Bradt H.~V., Remillard R.~A., Tuohy I.~R., Hoshi R., Nakagawa M., 1986, 
ApJ, 308, 635 

\bibitem[\protect\citeauthoryear{Macklin}{1982}]{macklin82mnras} 
Macklin J.~T., 1982, MNRAS, 199, 1119 

\bibitem[\protect\citeauthoryear{McClintock \& Remillard}{2003}]{mcclintock06} McClintock J., Remillard R., 2006, Cambridge Astrophysics Series, No. 39: Compact Stellar X-Ray Sources. Cambridge, Cambridge Univ. Press, p. 157 

\bibitem[\protect\citeauthoryear{Merloni, Fabian, \& Ross}{2000}]{merloni00mnras} Merloni A., Fabian A.~C., Ross R.~R., 2000, MNRAS, 313, 193 

\bibitem[\protect\citeauthoryear{Motta et al.}{2012}]{motta12arXiv1209.0327} 
Motta S., Homan J., Mu{\~n}oz-Darias T., Casella P., Belloni T.~M., 
Hiemstra B., M{\`e}ndez M., 2012, MNRAS, 427, 595 

\bibitem[\protect\citeauthoryear{Nowak}{2000}]{nowak00mnras} Nowak 
M.~A., 2000, MNRAS, 318, 361 

\bibitem[\protect\citeauthoryear{Oosterbroek et al.}{1998}]{oosterbroek98aa} Oosterbroek T., Parmar A.~N., Kuulkers E., Belloni T., van der Klis M., Frontera F., Santangelo A., 1998, A\&A, 340, 431 

\bibitem[\protect\citeauthoryear{Parmar, Angelini, \& White}{1995}]{parmar95apjl} Parmar A.~N., Angelini L., White N.~E., 1995, ApJ, 452, L129 

\bibitem[\protect\citeauthoryear{Parmar, Stella, \& White}{1986}]{parmar86apj} Parmar A.~N., Stella L., White N.~E., 1986, ApJ, 304, 664 

\bibitem[\protect\citeauthoryear{Parmar et al.}{1997}]{parmar97aa} Parmar A.~N., Williams O.~R., Kuulkers E., Angelini L., White N.~E., 1997, A\&A, 319, 855 

\bibitem[\protect\citeauthoryear{Priedhorsky}{1986}]{priedhorsky86apss} Priedhorsky W., 1986, Ap\&SS, 126, 89 

\bibitem[\protect\citeauthoryear{Psaltis, Belloni, \& van der Klis}{1999}]{psaltis99apj} Psaltis D., Belloni T., van der Klis M., 1999, ApJ, 520, 262 

\bibitem[\protect\citeauthoryear{Rao et al.}{2000}]{rao00aa} Rao A.~R., Naik S., Vadawale S.~V., Chakrabarti S.~K., 2000, A\&A, 360, L25 

\bibitem[\protect\citeauthoryear{Shakura \& Sunyaev}{1973}]{shakura73aa} Shakura N.~I., Sunyaev R.~A., 1973, A\&A, 24, 337 

\bibitem[\protect\citeauthoryear{Shimura \& Takahara}{1995}]{shimura95apj} Shimura T., Takahara F., 1995, ApJ, 445, 780 

\bibitem[\protect\citeauthoryear{Steiner et al.}{2009}]{steiner09pasp} Steiner J.~F., Narayan R., McClintock J.~E., Ebisawa K., 2009, PASP, 121, 1279 

\bibitem[\protect\citeauthoryear{Tanaka \& Lewin}{1995}]{tanaka95} Tanaka Y., Lewin W.~H.~G., 1995, in X-Ray Binaries, W.H.G. Lewin, J. Van Paradijs \& E.~P.~J. van den Heuvel, eds,  Cambridge University Press, Cambridge, 126

\bibitem[\protect\citeauthoryear{Titarchuk}{1994}]{titarchuk94apj} Titarchuk L., 1994, ApJ, 434, 570 

\bibitem[\protect\citeauthoryear{Tomsick, Lapshov, \& Kaaret}{1998}]{tomsick98apj} Tomsick J.~A., Lapshov I., Kaaret P., 1998, ApJ, 494, 747 
\bibitem[\protect\citeauthoryear{Tomsick \& Kaaret}{2000}]{tomsick00apj} Tomsick J.~A., Kaaret P., 2000, ApJ, 537, 448 
\bibitem[\protect\citeauthoryear{Tomsick et al.}{2005}]{tomsick05apj} Tomsick J.~A., Corbel S., Goldwurm A., 
Kaaret P., 2005, ApJ, 630, 413 

\bibitem[\protect\citeauthoryear{Trigo et al.}{2013}]{trigo13nat} Trigo, M.~D., Miller-Jones, J.~C.,~A., Migliari, S., Broderick, J.,~W., Tzioumis, T., 2013, Nat, 504, 260

\bibitem[\protect\citeauthoryear{Trudolyubov, Borozdin, \& Priedhorsky}{2001}]{trudolyubov01mnras} Trudolyubov S.~P., Borozdin K.~N., Priedhorsky W.~C., 2001, MNRAS, 322, 309 



\bibitem[\protect\citeauthoryear{Wagoner}{2012}]{wagoner12apjl} Wagoner R.~V., 2012, ApJ, 752, L18 

\bibitem[\protect\citeauthoryear{Watarai, Mizuno, \& Mineshige}{2001}]{watarai01apj} Watarai K.-y., Mizuno T., Mineshige S., 2001, ApJ, 549, L77 

\bibitem[\protect\citeauthoryear{Watarai et al.}{2000}]{watarai00pasj} Watarai K.-y., Fukue J., Takeuchi M., 
Mineshige S., 2000, PASJ, 52, 133 

\bibitem[\protect\citeauthoryear{Wijnands, Homan, \& van der Klis}{1999}]{wijnands99apjl} Wijnands R., Homan J., van der Klis M., 1999, ApJ, 526, L33 

\bibitem[\protect\citeauthoryear{Zimmerman et al.}{2005}]{zimmerman05apj} Zimmerman E.~R., Narayan R., McClintock 
J.~E., Miller J.~M., 2005, ApJ, 618, 832 

\end{thebibliography}
\end{document}